# INTERACTING AGENTS IN SOCIAL NETWORKS:

# THE IDEA OF SELF AND INFLUENCE SPACES

Fariel Shafee

Physics Department
Princeton University
Princeton, NJ 08544
USA

*Abstract:*

We have study the evolution of social clusters, in an analogy with physical spin systems, and in detail show the importance of the concept of the "self" of each agent with quantifiable variable attributes. We investigate the effective influence space around each agent with respect to each attribute, which allows the cutoff of the Hamiltonian dictating the time evolution and suggest that equations similar to those in general relativity for geodesics in distorted space may be relevant in such a context too. We perform in a simple small-world toy system simulations with weight factors for different couplings between agents and their attributes and spin-type flips in either direction from consideration of a utility function, and observe chaotic, highly aperiodic behavior, with also the possibility of punctuated equilibrium-like phenomena. In a realistic large system, because of the very large number of parameters available, we suggest that it would probably almost always be necessary to reduce the problem to simpler systems with a manageable set of coupling matrices, using assumptions of fuzziness or symmetry or some other consideration.

# 1. INTRODUCTION

In classical economics, every agent behaves rationally to maximize his utility curves. However, in an imperfect world, not everyone looks rational. Phenomena like war and spite appear very contrary to what a rational person would do. Some behavioral incidents look self destructive, and hence irrational.

The reason human interactions are possible, and it is also possible to have a coherent system of communication and trade, is explained by the idea of a common set of needs and agreements about the state of the environment shared by most people. In our model, we argue that the majority of people would concur upon a set of objects considered as needs most of the time, and would each try to maximize the satisfaction of those needs, while interacting with each other, and also with the environment with which they are independently connected. They would agree on the state of the environment on an average, and can communicate with each other regarding a mutually agreed upon idea of the environment. In short, we agree about the idea of a macroscopically entangled "environment" with which the agents are connected.

Classical economics explores this set of needs and formulates equations for every possible deal and interactions. However, it fails to account for the so-called irrational human behaviors. If a set of data observed by most people, and possible to be experimentally verified by the majority, did not exist, social networks would perish. However, together with the network with a common set of knowledge, there exist phenomena like revolutions, hatred, riots, and many basic instincts not explainable by a concrete set of knowledge. Attempts have been made to justify irrational behaviors as weak, inconsistent or culturally inferior and evolution-wise unfit.

However, in this paper, we try to understand the fine structure of human perceptions and needs to model possible chaotic or annealed behaviors of social lattices from a spin glass point of view (Edwards Anderson 1975; Sherrington Kirkpatrick 1975) without any a priori bias towards a certain logic system successful in adapting with nature at a certain time point.

Previously, we have tried explaining the unquantifiable human needs (Shafee 2002) and the variations in utility curves by taking Gödel's incompleteness theorem (Gödel 1992) into account. We can also justify our argument by considering Hume and relying on the skepticism about human experience (Hume 1777). Existing ideas in psychology try to describe how differences in perception may lead to different personalities (see e.g. Blake and Ramsey 1951).

Although by means of communications, agents agree upon a broadly shared idea of the world, each agent is connected to her world independently, and gains information about the world by means of perceptions and communications. For example, low levels of latent inhibition to the environment may be associated with both the concept of genius or madness in two extreme cases (Peterson, Carson 2000). The idea suggests that some people are prone to be receptive to more stimuli from the environment and hence be more connected to the environment.

. In Appendix A, we discuss abstractly how agents with overlapping perceptions do not necessarily have to see the exact world, but might be able to interact in a manner as to agree upon enough information to communicate and consider another agent similar to himself. However, as we develop later, the idea of similarity may be dynamic, and also a fractal type quantity.

We argue that the interaction among agents is a game of incomplete information, each agent only being able to retrieve partially the information the other agent possesses while guessing his next move. The information trapped in different agents may not always be transferable to another completely in an understandable way. As we prove later, some of the pieces of information and preferences may be conflicting, impossible to exist together in a consistent logic system and some times genetically determined and hence almost permanently local to an individual agent.

We have also argued (Shafee 2002) that although each agent is rational in the sense that he tries to maximize his needs, the needs themselves are not rational and are allowed to vary from agent to agent.

## 2. THE DEFINITION OF THE "SELF"

Although existing literature has tried to find self-similarity and relatedness among agents (Pepper 2002, Hamilton 1964, Connolly and Martlew 1999), the models are often simplified and static. Models exist to calculate similarities among genetically related agents (Dawkins 2006, Pepper 2000). However, recent mappings of genomes have shown that human genomes are 99.9 percent similar (International Human Genome Sequencing Consortium 2001) and members of two isolated and different clusters may have more overall genetic similarity than members of the same cluster [e.g. broad genetic difference in blood groups within the same group of people]. However, despite this statistics, evidence of altruism and affiliation is often seen within ethnic, social and cultural groups (e.g. see Bowles and Gintis 2004), and specific minuscule differences in genetic structure can sometimes get expressed with large consequences in interactions with the environment and other agents, while large segments of genetic material remain dormant. Examples can be given as follows: single mutations in one gene may lead to deadly diseases like Huntington's disease and small mutation differences may result in the broad expression of the type of melanocyte. However, a large percent of DNA remains as non-coding DNA or junk DNA in eukaryotic cells that do not trigger any action. Hence, the expression and weight of variables in an agent is certainly not a linear function of the genetic material but a complex weighted set of expressions and cultural norms that are not functions of the genetic program, but marks left by interactions with other agents and the environment.

The structures of social clusters and graphs have become a field of study recently (see e.g. White and Reichardt 2007, Gastner and Newman 2006, Newman et al 2001, Barbasi and Reka 1999 ) However, detailed analysis into the psychological and social causes leading to the fine structures of the clustering has not been studied. Hence, the study of social networks and the field of psychology and self-similarity leading to clustering remain disjoint.

In some papers (Bowles 2004, Gintis 2003b), the so called group or pro-social emotions are mentioned, but the findings related to group emotions in non-genetically related clusters remain restricted to observation and calculation of economic group stability assuming that those emotions as preprogrammed (Bowles 2004) or are parts of reciprocity between individuals (Bowles and Gintis 2003). In this paper, we model a unified measure of self-similarity among agents that takes into account an array of variables of all possible sorts. We extend the existing ideas by posing further observations about interacting agents:

> **The notion of self or identity is not static**: Extreme cases of the same person undergoing drastic personality changes can be observed after brain lesions or injury of the brain due to accidents (e.g. see Raitu 2004**).** Other physiological changes due to hormones etc. may also cause personality shifts (e.g. see **(**Schulkin 1999)**).** From everyday experience we can feel how after some time and experience gap the same person appears "changed", although a broad group of tags may indicate this is the same person as the one known before. Hence, the idea of "same person" is often associated with a continuity accompanied with changes. The ideas of continuity and changes within the person might be validated by his experiences where the past rationally precedes the present. However, to another person observing the first person with "gaps of time and space" between the two snapshots of the "same person" the change might not appear validated or rational, as he is unable to connect with the experience local to the first person.
>
> **Expression of Variables** Broad differences in self-similarity based clusters and also finer differences among agents within clusters indicate that there are some constraints and choices in how many variables can be expressed at each interaction stage or level (Shafee 2007)
>
> **Interconnectedness** of different variables and states defining persons are sometimes complementary, and some times contradictory, when placed together or allowed to interact, and the sharing of resources and the environment come into play.

**Changing Priorities**: The priorities or weights assigned to a preference or piece of knowledge by an agent is important and sometimes modifiable.

**Interactions and updates:** The interactions among different types of traits with various degrees of stiffness and their interrelatedness are important and often cause interesting dynamics (see Shafee and Steingo 2008 and Shafee 2004 for an example and detailed discussion).

We begin with some basic definitions and the relevant equation. The Hamiltonian of a dynamical system controls how the system evolves in time, the speed of the changes and the direction, for each identifiable component of the system that has some freedom to evolve in a different way from other components.

In the spin-glass system in physics the spins $s_i$ of the components $i$ are usually of the same magnitude, but with different directions, indicated by the vector sign. When the components of the vectors are expressed explicitly, the symbol we use is $s^a_i$ When two spins have the same direction, i.e. are parallel, they are usually attracted to each other, i.e. the Hamiltonian or energy is minimized. In our context the spins may represent the personal preferences of individuals, so that two individuals having same likings like to form a bondage. However, unlike a ferromagnet, where all the spins may evolve into the same direction forming a strong magnet where the strengths of most of the units are added together, in a spin-glass the coupling $J^{ab}_{ij}$ between different attributes $a, b$ etc. of different agents $i,j$ etc. may have couplings with different magnitudes and signs. Hence, it may be impossible to align all units, i.e. we have frustration. In addition to the interaction of the agents among themselves represented by the quadratic form involving the coupling $J$, and two spins, there may also be an external environmental field, say $h$, with which every agent interacts. Hence the simplest Hamiltonian for such a system would be:

$$H = J_{ij}^{ab} \; \mathbf{s}_i^a \cdot \mathbf{s}_j^b \; + \; \mathbf{h}^a \cdot \mathbf{s}_i^a \tag{2.1}$$

In the rest of the paper, we will be modifying this basic spin glass equation by taking properties of social interactions into account and form a master equation eventually. The interaction Hamiltonian H is similar to an anisotropic Heisenberg Hamiltonian in the sense that we have two arrays with different components (variables in different states) interacting and the minimization of total energy in an energy landscape is the asymptotic goal. However, as the number of variables and states are increased, numerous local minima arise and fluctuations (noise) and sudden changes (including sharp changes in the environment and mutations) contribute to the instability of the network in a local minimum. We call it an *anisotropic* Heisenberg model because the components of the vector do not have the same weights in coupling together.

In the rest of the paper, we will observe how semi-stable structures (groups, cultures, friendships) arise in the landscape where the variables with respect to which these boundaries are drawn, are also allowed to change. We will relate each of these semi-stable structures with an extended idea of "self" that is defined. The permeability of the walls between these structures is also observed and discussed in the light of interactions and some known models of psychology.

The types of interaction can be energy minimizing when we have two similar preferences placed side by side or two complementary pieces of knowledge or skills placed next to one another. However, the complementary states bond to support or maximize the individual agents' preferences. How bonds between skills are formed and are saturated has been discussed in detail. (Shafee and Steingo 2008).

The main notion of our model, that sets it apart from models of classical economics and game theory, is that the idea of self, or an identity, will be a rather slowly weakening concept of the self- similar variables. Each of the arrays mentioned earlier is defined as an indivisible unit instead of being an undefined object contained within a body or brain. The idea of self-similarity derives from relating with similar variables in the same (or as in some cases complementary, as explained later) state in other agents' arrays and forming associations. We use the term slowly weakening, because the idea of "self" decreases as another array contains fewer variables in common, and also the strength of the "bonds" as a function of physical or influence distance. By influence-distance we mean the placement of a variable in a position where it can be influenced by the agent or from where it can influence the agent. This distance is not necessarily solely a geographic distance. The idea of an influence space is developed further later in this paper (see section 7). As a result, our model of "self" is a fractal like idea with a basic array of tightly bonded or entangled weighted variables forming the shortest unit.

However, as the structures move up the levels, the entities belonging to one level feel the effects of lower and higher levels in different degrees and act as members of higher levels together with interaction influences within the other entities of the same level. A specific example is given (Shafee and Steingo 2008) when an agent acts according to his own preferences given the constraints that higher levels (e.g. social clusters) are also interacting among themselves. The effects of different levels reflected in an individual shift as the different levels also change interactions among themselves dynamically (see Shafee and Steingo 2008 for a detailed description of this type of dynamics).

In this paper, we concentrate on the individual and interactions among individuals. The effect of our model at a more "macro-economic" scale is studied elsewhere (Shafee and Steingo 2008).

We allow individual agents modeled as arrays to interact in a social cluster. As we include more and more self-similar variables from other agents placed in a cluster but also containing dissimilar variables, the idea of self weakens.

The interesting points about this array are:

1. The variables in the array are interconnected with certain weights. The weights construe how correlated one variable is with another. A similar idea with axiom network was proposed (Shafee 2005).

1a. The same variables may carry different weights at different interaction levels (see Shafee 2007). For example, an agent's blood group may be important in the internal dynamics inside the body and in rejecting foreign cells but has little expression or weight in a typical interaction among different agents. However, in our social model, we discard cellular levels and start with the basic array of an individual that can be taken as a unit of complex identity.

2. In the spirit of condensed matter physics, where a spin-1/2 object can come in two states, we say that each variable may exist is one of several possible states. For simplicity, we only consider discrete states in this paper, and avoid any discussion of weighted superposed states (which can be formed in the same manner wave functions may be constructed in quantum mechanics).

3. Only certain variables are expressed at a time, although the rest of the array remains attached, but unexpressed as coarse-grained approximations take place. The finiteness of time and energy and finite allocation of the two required for both the realization, and in some cases fulfilling of a variable, is responsible for the truncation of the array.

Networks arise partly because agents choose to associate with others who are similar to themselves in some significant respect (Lazarsfeld Merton 1944). The noticeable structure of the model is the individual's array existing as the connected smallest unit so that the variables in that array are attached, and when self-similarity is measured by an agent, calculations are carried out by adding similarities between all expressed variables. Hence, a second agent may have similarity in one variable but dissimilarity in another, if the first variable exists in the same state in the two agents and the second has conflicting states. The similarity assumed between two agents is a measure of a fuzzy approximation based on this series calculation, which is expressed as an expression of connectedness or difference.

In a related paper (Shafee and Steingo 2008) the concept of similarity was developed in terms of belongings in clusters and subclusters and by forming a series of weighted memberships in all possible overlapping topological coverings. Here, we view the idea of similarity more in a "microeconomics" sense in the spirit of the objective of this paper. We concentrate on the idea of similarity between two agents as they are allowed to interact with respect to the environment. We use knowledge from pattern recognition to create ideas of similarity and a fuzzy sense of "liking" or "disliking."

These arrays representing a basic unit of identity or self can be constructed or realized through interactions with the environment. In that regard, a basic unit of an individual can be seen as a semiclosed structure (Shafee 2007) that interacts with the environment in its attempt to attain stability and gain closure. However, as it interacts with the environment, it tries to align along preferred directions as directed by preferences; the identity of a second agent is reflected in the second agent's attempt to align the environment in *his* preferred direction. The second agent himself is perceived not as an array but as a part of the environment, which is a sum of atoms etc. The actions of the second agent is part of the environment that the first agent is also connected to and the attempt of each agent to attain stability is perceived through interactions with the environment that the other agent also realizes to gain insights about the other's array or identity. It is the changes or realignments of the agents' environment that allow one agent to guess the states of the variables of the other agent. Hence, the perception of the second agent's array is "felt" through how the struggle to align the same environment in contradictory directions or in the similar direction affects one agent's time, energy and stability. Theories about how an agent perceives his own traits differently from the traits of others exist in current psychology literature (see Klein et al 2004, for example). Our abstract way of representing trait arrays for different agents, so that the traits of others may be perceived in an indirect way, is consistent with psychological findings.

This extended fuzzy idea of "self" may create some loosely or strongly defined clusters where the agents relate with self-similar structures in specific variables. The affinity with agents may also be a function of time, as which variables an agent would tend to find more important may be a dynamic function of time.

When an agent invests in perpetuating "self", we thus add a correction term to the self that includes other agents possessing similar variables. However, granted that no two agents are similar in every possible variable, any investment in perpetuating other agents with similar variables also comes with the cost of perpetuating some dissimilar variables.

These affinity clusters may be modeled as follows:

1. An agent perceives an extended idea of "self" leading to the creation of affinity clusters. This idea of "self" comes from relating to other agents who have similar variables in the same state as those of the agent himself. These variables are weighed. Hence, another agent who has a variable that is highly weighted in the same state as the agent's own is found to be "closer" or more related to the agent himself.

1 a. What is interesting is that the agents may or may not have knowledge of the proper state of these variables. Therefore, each variable in an agent has an actual value, and also a value perceived by the other agents.

2. The individual utility curve can get distorted because of this "*affinity factor*" as the corrected utility would be the utility of the individual corrected by a weighted utility of the affinity group. The affinity factor will reflect an agent's perception of other agents' variable values.

Our model is built from the idea that "self" is propagated, and perpetuated with preferences and variables related with "self" maximized. When the notion of self is used, the notion of a fractal type weakening structure is used, with centers existing as individual arrays, so that as first degree approximations, we retrieve an individual and the clean laws of individualism, which, however, is modified, as second and higher order interaction terms are no longer neglected given certain situations. If we write an equation in terms of an expanding series where the first term is based on the agent's own intrinsic array of preferences (or values) and the latter terms derive from his grading of interacting agents or his realization of agents who are similar/dissimilar, we obtain an expanding series.

This series is similar to an infinite Taylor series expansion. This is normally cut off after taking into account a certain number of terms depending on the context, since higher terms are usually more and more negligible. The first term accounts only for the individual and ignores all similarity/interaction effects so that the identity created using the first term only retrieves the basic rational being.

How higher order terms come into play within any network and interaction is dependent on the value of the higher order terms and may differ from one scenario to another based on total number of agents and the partitioning of similarities and the weights assigned to the variables at a specific time and place. Hence, although in some cases the basic laws of classical economics can be retrieved and applied very easily when higher order terms can be neglected, in many other cases higher order terms are observed clearly and in instances (see section 9.3) certain higher order term may overshadow the first order term and lead to perplexing actions from an individual.

$$U = U_o + \alpha U_1 + \alpha^1 U_2 + \alpha^2 U_3 + \ldots \tag{2.2}$$

Here U is the utility. The first term is the utility of an isolated individual agent, while the second term is the utility from the second level affiliation of the individual with one other agent, the third being the result of interactions among three, and so on.

It is interesting to note that the term at each order may include influences from conflicting or complementary groupings of the same order with comparable amounts of similarity but conflicting dissimilarities.

A miscalculation or bluffing on other agents' parts will cause the agent to invest energy in perpetuating dissimilar or contradictory variables. However, each member within a known affinity group will have the individual affinity and the corrected group affinity both playing a role in the utility curve. Moreover, the group utility is a factor that is shared by all members of the group, and each individual will tend to utilize the other agent's affinity points to gain an increase in utility at no or little expense of ones own cost of entropy. As a result, it might be profitable to "*net utility*" to invest in increasing other agents' affinity utility factor.

Again, investments made in maximizing group affinity utilities from others will result in an expense within a closed sub domain. If the total amount of investment or energy to be spent is kept constant, an increased investment within the sub domain will reduce investments in games with other sub domains and with nature. As a result, agents in a sub domain playing against each other with a large weight factor to increase each other's group affinity factor will have less to invest in games outside the sub domain. Again, bluffing about ones variables might be a strategy an agent uses to gain group affinity points from agents with dissimilar variables with no expenditure of group affinity from the agent's own side. We can check out by simulation cases where agents indeed sharing affinity variables are put together in a cluster, and clusters where some concealing agents are mixed.

The other interesting property of this notion of "self" is that the variables in the unit array are dynamic in the sense that with the shift of weights (priorities) and given the constraints posed by time and energy, variables expressed within an agent might change.

As a result, the idea of self, or an identity will change with time. However, when a certain variable is connected to the environment and there is a sudden shift in the environment, the agent's variable(s) connected to that environment also need to undergo shifts.

As the basic array gets modified, so does a person's perception of "self" and the affinity clusters that appear as weaker versions of "self" also redraw their boundaries.

# 3. VARIABLES:

## 3.1 Knowledge, Preference Variables

Knowledge is a set of information that an agent obtains by interacting with nature and other agents. Knowledge is obtained by symmetry breaking and provides an agent with a rule or a measurement that is in a collapsed state and can be used in the future to play against nature (Collier 1996). A knowledge or piece of information can be used to satisfy the agent's preferences of what should be done in the future. Although knowledge can be shared among agents, so that agents can decide on a possible set of largely overlapping common knowledge, the preferences about how these knowledge pieces should be used to interact with nature is not necessarily a common or correct set. Preferences, or utilities, may vary in importance or weight from agent to agent.

We can describe the situation as follows. We can say that the future is a superposition of many possibilities and can collapse to only one of many possible states to which all the agents will be connected. Now, different agents will bid for different futures. However, together, they will all collapse to only one of the possibilities involving the set.

We have modeled this type of collapse at a quantum level (Shafee 2005b) by allowing the possible states to go through a first passage random walk. Our work, however did not involve any quantum postulate, but depended on randomness of a classical kind. We can imagine something very similar taking place here, only at a macroscopic level, the agents' different preferences bidding for one of many possible futures.

Resources can be explained as specific sites of interactions with nature can be used to change nature in the direction of an agent's preference. Resources are finite, and the same piece may be needed for agents with conflicting visions.

Each of the agents would consider their vision to be necessary to some degree for their own perpetuation or existence, and hence would compete with others for that resource.

## 3.2 Skill and Aptitude Variables

Inherent talent or aptitude and training both come into play in the final skill of an agent. Although extreme cases prevail: some people cannot be trained in certain skills even to the level of an "average" whereas some people like geniuses may excel (e.g. see Johnson 2004) and produce at the top level with little or no training, for an average person a combination of ability and training are important.

Training can be defined as follows: The interactions of the agents with nature in order to modify nature depend on both skill and aptitude. An agent can modify his ability to interact with nature by repeatedly interacting with nature. It is a mechanism of adaptation by which the agent learns to optimize his efforts to interact in a certain way. However, the innate ability of an agent to interact in that certain way or perform a certain deed can be contributed as talent or aptitude.

An agent with less aptitude will need to invest a higher amount of time and energy to master a certain task. We can model aptitude as follows: We can assume that each agent is born with a group of preference curves that are slightly different from other agents' preference curves. The curves can be modified by interacting with nature, and can be brought to resemble roughly another person's preference curve distribution. However, adaptation and learning will create an exact match with another distribution very rarely with the increase in number of variables and steps in learning process involved with a particular distribution. A specific distribution, again, can be optimized to interact with nature in specific ways and produce specific results. A specific agent's aptitude for a job may also be interrelated with his or her aptitude for other jobs, as performing a certain group task optimally may require sharing one or more preference variables. However, these fine tunings will rarely match to an exact optimized result, and will involve a probabilistic and stochastic element involving all agents. A more physics based model of hierarchy in skills has been developed (Shafee and Steingo 2008)

### 3.3 Beliefs and Faith and their Strengths

An agent acquires a piece of information by interacting with the environment by means of his perceptions and can communicate the information to other agents who can also independently obtain the same information by interacting with nature. The information has a credibility factor assigned to it taking the following factors into consideration:

**A.** Whether the information can be verified by the agent interacting directly with the environment, and by using simple logic starting from the basic axioms the majority of agents agree with.

**B.** If the information is obtained indirectly by communicating with other agents instead of interacting with the environment, the frequency of the information received, and the credibility of the source

**C.** Whether the information contradicts other information held by the agent as true, and the degree of credibility of the already held conflicting information

When the information is received by interacting with others, and cannot be directly verified by interacting with nature, that is more a belief.

# 4. VARIABLES AT PLAY

## 4.1 Self and Mutability with Respect to Fitness

An agent will try to perpetuate him-"self" and in order to do so, he will play against nature and against other agents. We tried formulating this game in our first paper (Shafee 2002). However, while playing against nature, a certain set of variables will be fitter at a certain time-space point than another. An agent will try to dynamically change the idea of "self" to make himself fittest to perpetuate. In other words, an agent will either change variables to be placed optimally with nature, or change nature to fit him best.

As a second strategy, an agent will try to include other agents with optimal variables in the cluster of "extended self". This can be done so by mating with an agent with the coveted variable to include the variable in the agent's own array.

## 4.2 A possible scenario for Beliefs, Betting, low probability high stakes and the splitting of clusters

The existence of these arrays consisting preferences or axioms (Shafee 2004) must include an existence axiom in common. This is by tautology that an agent not wanting to exist or not wanting similar variables to exist will simply not exist.

However, the experience of death and perishing and the inevitability of so may act as a negative feedback and cause the destruction of these complex arrays.

We argue that the existence of spurious axioms or beliefs is necessary and evolution-wise stable especially when high stake situations arise or when societies/individuals reach crucial points where survival chances are low.

Let us say that a social cluster reaches a point when the only mode of survival is a high-risk action with a survival chance of .1 percent and a possible high reward if that high stake bet comes out in favor.

A rational person in a social cluster connected with other agents will weigh the possibility of other agents carrying out the high-risk job from which the entire cluster might benefit. However, each agent would have very little incentive of doing it himself and as a result, only few agents who are incapable of calculating the risk properly will undertake such high stake tasks. The statistical outcome of such a scenario will be very much against survival.

However, in such a critical juncture, if spurious axioms or beliefs are inserted that reduce the risk factor, more agents will be inclined to undertake the activity and hence the statistical outcome of betting will have more of a chance of having one or more successful drawings.

Let us say, that in a juncture of this type, a belief is invented that a spurious reward exists that is added to the possibility of death and hence increasing the total weighted reward. If this belief can be propagated within the social network, more agents would undertake the risky activity and the statistical outcome would be beneficial for the network. Beliefs of this type can be propagated in the same manner rumors are

propagated (Moreno et al 2004)

The partitioning of a network based on the propagation of such beliefs and the success of one subgroup in attaining a high-risk success is an interesting problem, especially after these beliefs attain higher weights as successes are correlated with them although the bases of these beliefs themselves might be spurious. As the "believers" who took a higher risk than the unbelievers, an agent from the "believer" partition would be less likely to share his success form a member from a "non-believer" group and the partitioning may become sharper with the beliefs attaining higher weights as more agents place bets with high risks using those sets of beliefs.

However, as points of these critical junctures are passed, the beliefs that had attained high weights may remain is such position as part of the agent's identity and may lead to non-optimum scenarios for the cluster or the subgroups at points although these sets of belief once helped these groups to survive. However, sudden removals of these beliefs following critical junctures are inhibited because:

1. The success of the beliefs at the critical points depends on how strongly an agent can be made to believe them. Hence, the stronger a set of beliefs, the higher success they will have at a critical point and the tougher they will be to remove.

2. A sudden disbelief in a set of beliefs is prohibited by inertia and continuity of identity.

3. A sharp partition in a social cluster based on a belief will have numerous other variables related with the partition after the partition has been made including group utilities and group risks restricted within the partitions.

4. The continuation of the belief acts as a hope that the set will help a second point of juncture (by means of experimental induction).

## 4.3 Correlated and Uncorrelated Variables and their Interpretations

Variables may be obtained by genetic inheritance or by interactions with nature and with other agents. Again, the variables an agent possesses may or may not be correlated.

Moreover, they may appear to be correlated if they are placed in a certain environment for a certain time period and then become uncorrelated when the agent is removed from the environment. For example, if placed in a certain environment that requires optimization of two variables for survival, an agent may develop specializations in two factors. However, with the removal of the agent from the environment it is possible for one of the variables to become randomized while the other remains fixed. Again, some variables may be dominant and others recessive. As a result, mating among specialized agents and non-specialized agents may produce offspring that are specialized in one variable but not specialized in the other.

Let us consider two networks of arrays with the partitioning factor of success in a certain skill or variable: A. Let us also say that another variable B is mutually exclusive so that B is part of "culture" or tag of the successful cluster where the second cluster is B'. The two clusters are sufficiently separated so that only easily discernible variables in macro scale are observable from to members of one cluster to another. Now, let us suppose that several connected but minor variables lead to the success of A in cluster 1 which is not observable from cluster 2. Alternatively, let us assume that cluster 1 succeeded in A simply as a matter of chance.

However, the mutual exclusivity of *B* and the lack of experience of the existence of B in cluster *1*, might lead to a possible selling point that *B* and *A* are correlated and the successful cluster might try to "export" *B* to cluster *1* who is *interested* in achieving success in *A*.

## 4.4 Optimized Variables and Bluffing

The agents use their set of variables to play against nature and also against the other agents competing for a finite resource offered by nature. Therefore, although an agent may be interested in utilizing another agent's more optimized variable, they are also playing against each other to maximize the perpetuation of their own variables. At a certain time point, an agent may not have a set of variables that are all optimal with respect to nature.

Agents can form a network where more optimized variables of one agent are used by another agent at that point. In return, the other more optimized variables of the second agent are also shared. However, the possession of more optimized variables puts an agent in a position with higher bargaining power. The game at this point can very easily be modeled in the same fashion as several agents with cards with higher or lower values. An agent may show the value of the card before placing a bid in the game or bluff. In a pioneering model on game theory (Neumann Morgenstern 1944), the authors argue how bluffing is an essential strategy of any such game. Stories of bluffing and passing as the member of one group while believing in another abound in history, especially during war-time (see e.g. Shannon and Blackman 2002). Situations with puppet kings can also be attributed to this (e.g. see Hugh 1969).

Similarly, in the game of social lattice, equilibrium points are achieved by bluffing and by attempting to interpret the correct optimizations of the variables possessed by other agents. The inclusion of an optimized variable in an agent's array by mating with an agent possessing that variable also includes variables previously not in that array and not optimized with respect to nature in the array of the agent.

This new array will form a new definition of self and will thus isolate the agent from its older cluster defining self if the newly introduced variables are dissimilar enough with the first set of variables. As a result, an agent not wishing to break his former "self" cluster might as well find strategies to make use of the other agent's optimized variable without making the second agent part of him "self".

# 5. VARIABLE DYNAMICS

## 5.1 Flipping of Variables: Critical Net Utility and Correlation among Variables

In a follow-up paper (Shafee 2005), we have made some simulations with our social spin model taking the following points into account:

1. a **coupling strength** $J^k_{ij}$ (where $i$ and $j$ denote the two agents and $k$ the variable), that takes into account the weight given to the certain variable by agent $i$, when interacting with agent $j$.
$J^k_{ij}$ is the same for all $j$ for a specific $i$, as the weight would depend specifically on agent $i$'s valuation,
2. the **number of agents** in the network, or the total number of nodes in the cluster carrying the same variable,
3. the **influence distance**. In this paper, we add an extra feature to the model, that is the influence distance between the agents, as described later.
4. **shifting states of a variable**: In an idea similar to quantum mechanics, but in an abstract macroscopic level, we assume that each variable can exist in more than one possible states.

For example, an agent may have an affinity for a certain color. One agent might find the color white peaceful while another agent might find blue more tranquil. Therefore, the two agents have two different states of the same variable. Extreme cases of these preferences may be expressed as addictions.

So, the state white has something similar to an energy minimum in one agent while the state blue might have an energy minimum in another. Let us call these states the ground states in the two agents. Now the two agents may have two different fixed ground states determined.

Some traits are known to be anatomically fixed in humans and can be attributed to genetically determined structural differences in the brain and appear as parts of ones personality (see e.g. Rauch et al 2005). Some preferences can be related to these personalities. In extreme cases of addiction, for example, it has been shown that the addiction for nicotine might have a correlation with aggressive personalities (see e.g. Fallon et al 2004)

If a ground state is genetically fixed, then the agent can be in another state only when he is in an excited state. This excited state may not be locally energetically favorable, but induced either by peer pressure or in order to be consistent with the other variables within the agent.

Alternately, the ground states might be defined by interactions and experience, otherwise known as an acquired taste. Extreme cases of cravings after addictions can be attributed to both physical and memory-based reasons (see e.g. Tavares et al 2005). If a ground state is reached, the agent may be able to flip to another state permanently at some cost. The one time cost would include relearning this agent's new taste and also the continued costs for satisfying the new taste.

Now let us say that the network as a whole can minimize total energy by allowing for locally excited states. So in order to have a connected network in an optimal state in the lowest energy state, all agents need not be at their own minimum energy state. This might happen if a genetically fixed ground state needs to be excited to a non-preferable state to minimize interaction energy.

However, these local excitations are subject to bargaining powers of the individual agents placed in the network and access to knowledge about each agent's true preferences and needs by other members.

We can draw analogies with equations from physics where we associate energy levels with preferences. External fields may be similar to light or laser pulses causing excitations to higher energy levels and a constant pulse might be required to maintain a higher energy level state. Using lasers coherent light can be used to shift energy states of electrons. Expectation values of variables in a mixed state can be manipulated by using lasers (e.g. see Lloyd 2000).

## 5.2 Time, Energy and Truncation

In a previous paper (Shafee 2007), we have developed the idea of semi-closed systems to relate with complex entities. Our idea of "self" is that semi closed system that also updates itself continually with some possible modifications while interacting with the external network.

By minimizing the Hamiltonian where the variables related with an agent interact with other agents and the environment, the agent increases its chance of stability, and hence tries to minimize its interaction energy. The sum over other agents may be truncated to only those which couple relatively strongly with at least one of the attributes of the agent [$i$] being considered, and one may use a convenient cut-off for this purpose. Even with a truncated set the actual renormalized interaction, taking into account interactions involving intermediaries, may be quite complicated and should be represented by a general form $f$

We can, therefore, rewrite the original spin glass equation (2.1) as

$$H_i = \sum_j f(s_i^a s_j^b) = \sum_j J_{ij}^{ab} \mathbf{s}_i^a \cdot \mathbf{s}_j^b + \mathbf{h}^a \cdot \mathbf{s}_i^a \qquad (5.1)$$

Here we assume that the function $f$ is additive. The complex entity exists as long as it is able to hold the self-array together with a low energy. Let us define the threshold to be zero so that as soon as $H_{self-interaction}$ becomes positive, the array variables break up and dissipate among the environment.

The terms of interaction with other agents and the environment come with cost factors of time and energy, and also with flipping costs. The finiteness of time may be seen as the variables constantly interacting with the environment in certain variables causing the self-interaction matrix to weaken, as there are larger degrees of freedom in the environment than within the self. Dissipation of energy is also seen as weakening of self-interaction bonds, which can be somehow compensated for by interacting with the environment as well and reconstructing those bonds to some extent.

However, the same variable may interact with different degrees of the environment, and the environment updates itself in one of the variables and not other inconsistencies may arise.

For example, an agent may be genetically adapted to a local environment in one variable and acquire a quality that is supported by several environmental variables. However, this fixed genetic state as a result of environmental interaction may pose to be inconsistent as another related environmental factor changes. We can say that the genetic mutation is fixed by constant interaction with a certain variable with the environment, which can become non-optimal for the agent as another variable of the environment that interacts with the mutation changes.

The truncation of the interaction array must occur in the agent's realization since the number of variables within the self interacting array performing the calculations must be smaller than the total number of variables in the complete Hamiltonian, and it is necessary to choose a finite number of terms to perceive from a continuous (in time) stream of terms received. A truncation of the array must occur where important terms in the series with large weights become significant and the rest remains suppressed.

We can thus rewrite the master equation as

$$H_{perceived} = J'^{ab}_{ij} \text{(perceived)} \, \mathbf{s}_i^a \cdot \mathbf{s}_j^b + \mathbf{h}^a \cdot \mathbf{s}_i^a \qquad (5.2)$$

with the constraint that
$H_{perceived}$ = ordered array truncated when storage of perception at a time = 0,
and the assumed unchanged variables gain a low weight in $H_{perceived}$ as efficient storage would imply that assumed steady terms are calculated less frequently.
$H_{perceived}$ fixes the weight of a variable in the agent's perception and is the weight used by the agent when he chooses to interact with the environment and other agents. Hence, a variable that is vital for the agent's existence may appear, at one point, to be unimportant because it is perceived as stable.

The agent tries to minimize the perceived Hamiltonian while the actual Hamiltonian that is affected by his actions might have terms he had put zero weight to because of truncation.

## 5.3 The Intergeneration Journey of the "difficult to flip" Variables

The idea of perpetuating "self" of identities was discussed (Shafee 2002) as a tautological clause in the game, since any identity not interested in perpetuation will simply not be perpetuated. We expanded the idea to add a broad idea of self. However, this identity, besides forming bonds clusters with other self-similar agents, would also exist temporarily. However, the as a complex autonomous agent (the unit array) strives to achieve some sort of stability against nature that vies to increase entropy, the game of information and entropy is played (Shafee 2002) where the agent gathers information to achieve stability and the interaction with nature to gain the information eventually leads to increased entropy and decay.

However, the agent's perpetuation of self is reflected in the creation of new generation of agents with genetic mingling: to reduce the effects of harmful genes being carried on with particular traits and also to propagate his own genetic material. Genetic algorithms (Schwefel 1974) based on these principles show how such a mechanism is favorable in an energy landscape (perhaps only locally in an environment that undergoes change) to produce structures that are more optimal with respect to the environment with each agents' chance of passing some fraction of his own genes to that structure. As an agent passes his genetic material to another generation, it is the small genetic fragments giving rise to broad appearances that get passed on with a certain probability, while broader bundles of genes giving rise to more intricate mechanisms like IQ become isolated. Hence, with each generation, skill types can change considerably while maintaining a group of genetic markers.

An agent's urge to propagate his complete identity in time is often reflected in educating progeny in similar tastes and norms. The vertical propagation of norms (Gintis 2003a) is often the most influential way of propagating values and norms across generations. The urge to pass one's complete identity to the next generation is challenged by the reshuffled and shifted genetic preferences of the next generation. However, the trust in vertical propagation of norms is required to take the constraint of time and energy into account in acquiring information so that each agent is incapable of forming his own complete set of values and rationales. The vertical propagation of norms is trusted because of the already high degree of correlation of genetic material in the identity that makes the set already favorably bonded so that the members only increase in the perpetuation of self-similarity in genetic material by acquiring norms by trust. Hence, it is a type of feedback that creates the balancing attractive force against the reshuffled identity.

The shuffling and the stiffness of the discontinuity of generations were discussed to some extent. (Shafee and Steingo 2008). We mention a few more aspects of the mechanism. Interestingly, the intergeneration shuffling is related to the propagation of small fractions of genes used as markers. This shuffling leads to the lack of correlation of those markers with broader genetic bundles and also acquired tastes, and how the skill of an individual accompanied by his identity are played against the skills of slightly different agents and the effects of markers and correlations. For example, the similarity of the appearance of eyes in parents and children can be easily detected and correlated, but less correlation in the propagation of skills is notable. Attempts to correlate jobs with lineage, as is practiced in some places in India by the creation of a caste system, often lead to disastrous results (Iyer 2007).

## 6. WEIGHT FACTORS OF VARIABLES

We describe each agent as an array of an infinite number of variables. We consider cases where a finite number of variables carry a changeable but large portion of the total weights. On the other hand, if the weight factors were arranged thinly among many variables, which are uncorrelated among agents, no clustering would occur. However, we consider the weight factors to be also dynamic. Let us consider $n$ leading variables among $m$ agents.

Let us say that agent $i$ has a weight for variable $k$ as $W_{ik}$. In that case agent $i$ will also invest $W_{ik}$ proportion of its energy in optimizing in variable $k$. Also, agent $i$ will form positive or negative bonds with other agents which will contribute to the agent's "affinity utility" as a function of $W_{ik}$. However, if agent $j$ has a weight $W_{ik}$ for variable $k$, agent $k$ will gain from agent $j$'s affinity utility as a function of $W_{ik}$. In addition, flipping a variable $k$ will affect an agent as a function of $W_{ik}$ and $\sum W_{ik}$ and the alignment of $k$ in other agents.

The weights come into play significantly in the following way: The agent can only afford a limited amount of change in entropy, or in other words, the agent has only a certain amount of time and energy available for spending. This constraint will be taken into account when net utilities are maximized. The agent will start at optimizing the highest weighted utility, or existence, and will go down the tree by optimizing utilities that are connected to "self" by taking the weight factor and the net utility into account. The variable to observe is the utility scaled by the weight factor with the entropy factor subtracted. When several nodes are reached from one node that represent the same net utility within a certain error range, with the weighted utility put in, the nodes are pursued in parallel as we go down the utility tree.

## 6.1 Weight Factors, Risk Factors and Integration over Time

We assume that agents placed in a social lattice will play against one another and against nature to optimize their utilities. However, with every game, we can associate a risk factor. For example, if several agents are placed in a market, and each of them values two different commodities differently, each of the agents will try to deduce the other agent's valuation in order to maximize his/her own profits in the futures market. This valuation can be guessed if enough information is collected about the second agent's past decisions (see theories about games of incomplete information Fudenberg and Tirole 1993). However, the utility curve of the second agent is also subject to change. As each of the agents interacts with the environment separately, they acquire more and more information, and their needs may reflect a changed set of information possessed by them. The importance of a certain utility may also go down or up as new information is added to an agent's information system. This possibility of change can be lumped into a risk factor.

How a certain agent calculates this risk factor also affects his/her decisions. Again, calculating the risk factor or possible future actions requires an investment of time and energy. Since each agent tends to minimize the energy spent, how much energy an agent will invest in interpreting the second agent's future actions will also depend on the first agent's interpretation of the "importance" of the second agent's actions.

The weight factors are very similar to diversifying ones portfolio; an estimation of investments made into different utility-stocks with long term and short term options, similar to hedging (see Black et al 1973). Utilities will be connected with weight factors that will be proportional to the risk factor associated with the certain utility. Besides, possible changes are taken into consideration when integrating all utilities over time. In a many step game, the expected payoff from the n*th* step depends on the on integrating over all the risk factors over time. The weight factor will also depend on the possibility of the maturity of an **n** step game. The other term to be taken into consideration is the possibility that the utility variable that the game is optimizing on will not flip by the maturity of the n-step game, as with a flip, the payoff from the game will become negative. Other minor terms to be taken into consideration are possible inclusions of agents in strong affinity clusters that will distort the utility curve and shift the value of the payoff relative to the agent. The existence axiom must be the most highly weighted variable, which we assume to be fixed. The existence axiom, as defined in section 4.2, can be described in detail as an agent's utility in perpetuating the array of weighted variables in the closest possible unchanged form so that the highest possible utilities are obtained from the variable, taking the weights into account. However, when the utility fall below a threshold, a variable can be updated, as the utility is not contributing to the existence axiom then. If, taken the flipping energies into account, a flipped variable produces a higher than threshold utility, the flipped variable will redefine the definition of self. Again, some variables are connected to the self axiom with a high flipping energy threshold and also a difficult to modify large weight factor. These are the variables that connect the agents with the environment or nature in a material way, so that flipping them will inevitably cease the existence axiom. For example, eating and seeking shelter are utilities that are very difficult to flip, though the preference in eating might be somewhat modifiable. Therefore, some variables that are used for linking with nature have a high flipping energy. This is somewhat similar to a spin system being linked to a larger thermodynamics system where the variables are controlled more by the larger system's average than the individual fluctuations of a small system connected to it. The utility variables linked with the existence are again optimized because they are expected to perpetuate the existence. The existence axiom will also take into account the coupling strengths among the variables when defining the meaning of existence. For example, the variables inside an agent are inter-connected closely, and flipping one of them affects other variables in the agent's array strongly. However, the couplings with other agents' variables are long range, and an internal shift in an agent's variable will have a long-range effect in other agents' similar variables. As a result, the concept of self is concentrated most within the physical agent himself and fades away as longer and longer range couplings, and also couplings with agents with more and more dissimilar variables, are reached. Some of the weights might be easily shifted, whereas others might have a hard shifting possibility. For example, material needs, such as food and shelter have a high weight factor determined by nature. These weight factors depend on the agent's game with nature rather than the agent's interaction with other agents.

We can see how the notion of risk derives from the truncation of a series with variables and terms exceeding the calculating cost allocated for that certain action.

The necessity for approximating a series of this type is inherent with the idea of coarse graining involved with the production of structures of different levels. How only a finite number of variables are expressed in a level because of coarse graining has been discussed (Shafee 2007). This paper deals more with the approximation carried on by an individual and hence is similar to microeconomics with correction terms added because of interactions and self-similarity. In a recent paper (Shafee and Steingo 2008) macro behaviors that lead to the properties of level structures like entire clusters and interacting societies have been discussed in detail.

The act of truncating a series is necessitated by the fact that a finite amount of time and energy are available to each agent to perform a calculation and task and he is endowed with a finite amount of both. Hence, the diversification of time and energy (together let us call them cost factors) among multiple activities and tasks is required to avoid "infinite loops." The detection of important terms and properly minimizes risk factors but may come with the cost of locally higher costs. However, the residual terms form each interaction at a local level might be attached to interactions at another point of interaction or another calculation where they gain high weight and might appear suddenly as the interconnectedness of agents and the environment is reflected after multiple runs of interactions. This is somewhat similar to a model of quantum neural network (Shafee 2007b) where the effect of one qubit can be reflected in the state of another qubit that is not connected with it directly but by means of other intermediate components. Situations where small changes are later amplified into large bifurcations can be explained in the butterfly effect in chaos theory (see Devaney 2003; Holmes 1983).

## 6.2 Learning and Feedback in the Change of Weight factors, agreed Priorities and the Emergence of Cultures

A detailed model with weights and culture has been developed (Shafee and Steingo 2008).

The learning process, when placed in an interactive network where some of the weights are plastic while some others are genetically fixed may undergo changes leading to one of the following:

1. same variables with high priority but genetically fixed different ground states leading to conflicts;

2. same variables with high priority but different genetically fixed ground states (However, some variables excited to higher states artificially to achieve homogeneity);

3. all variables in their genetically fixed ground state, but with different priorities to reduce the effects of conflicts so that another agent being in a different state is minimized;

4. acquired taste variables with high priority.

The dynamics with a cluster will depend on the initial condition and the level of initial disparity. Specific case studies were performed (Shafee and Steingo

2008).

### 6.3 The learning process of the W factors:

In this section we mention some causes that mimic a society's evolution when agents are allowed to interact with one another and trade. In this section we take into account the simplest possible case where most of the priorities can be changed by allowing agents to interact with each other.

**A.** however, some weights are more strongly fixed by genetic makeup, and different agents may have slightly different fixed weights for the same variable. If the genetically fixed weights are shifted, the agents will have to pay a price for as long as the weights stay shifted.

In other words, the fixed weights have an equilibrium position, and an agent with the weights shifted will shift it back to the original equilibrium position when freed from peer pressure. The genetically fixed weights are the ones that are determined by the agent's long term or intergeneration.

**B.** Some other weights have a one time shifting cost associated with relearning the new weight and reorganizing other priorities. However, once the reorganization process is carried out, the shift will not cost the agent anything more, and the agent will keep the shifted weight when freed from peer pressure.

**C**. agents with different weights for the same variable who are allowed to interact with each other may have

i. the variables in the same state;
ii. the variables in conflicting states.

If the variables are in conflicting states, two highly weighted conflicting states interacting closely will be costly. However, when the variables are in the same state, if highly weighted, they can lead to the emergence of a group identity. Sudden rise of group identities after revolutions can be seen in history, e.g. the French Revolution, the Russian Revolution etc (see e.g. Doyle 1999; Bunyan and Fisher 1934).

## 7. INTERACTION AND INFLUENCE: DEFINING A NEW TOPOLOGICAL SPACE AND METRIC

When the interaction equation (2.1) was expressed, it was mentioned that interactions take place within a threshold influence distance. It was also mentioned that the coupling or interaction strength is dependent on how far away a second agent is placed from an agent in influence distance. An agent placed close to another person in the influence metric will have a larger effect on him and hence will have interaction higher weights.

We expand the idea of this influence metric here.

In an Ising model or spin model, interactions sometimes occur with only nearest neighbors and sometimes with equal strength (although signs may differ) among all members of a lattice. However, in our model, in order to accommodate for real social aspects, we introduce the idea of an influence space where, as is experienced in real life, interactions with every other member of a social cluster does not tend to have the same impact in a person's life.

In this section, we take interconnections among variables within agents and in nature into account to form the idea of a topological interaction space.

We first introduce two ideas. The first is an idea that is derived from the ideas of correlation and entanglement in quantum mechanics.

The distance metric to be defined in the influence space that will also account for nearest neighbors in the social network will incorporate the following properties:

1. The space need not be commutative. So $d_{ij} \neq d_{ji}$. This is because an agent with a higher influence will be able to reach many agents with fewer efforts than a non-influential agent. This could be an agent placed at the top of a work hierarchy pyramid (see Steingo 2007 for detailed construction) or in extreme cases it could be one person attracted to some quality of another obsessively, while the second remains non-responsive.

Therefore, in the influence metric, one agent might be very far from many agents to be affected by them, while those agents may remain close to him, and be easily affected by him.

2. If we assign influence coefficients between each pair of agents, then we can create a multiplication rule that is not commutative.

Say, the influence coefficient between A and B is 0.4 and that between B and C is 0.8. Then we can say that the influence coefficient between A and C is

$$\sum_B c_{AB} c_{BC} \tag{7.1}$$

This can be construed as how two apparently uncorrelated people can still have effects on each other's life because of a common connection, so that the action of one pair of interactions is reflected on the second pair.

## 7.1 Review of the Socio Metric

In a previous paper (Shafee 2005) we defined a metric $g^{ab}$ in a manner similar to the same variable found in general relativity.

In Euclidean space distance between two points **r₁** and **r₂** is given by

$$d(\mathbf{r_1},\mathbf{r_2}) = (x_1-x_2)^2 + (y_1-y_2)^2 + (z_1-z_2)^2 \tag{7.2}$$

In a general space the distance is given by

$$d(\mathbf{r_1},\mathbf{r_2}) = g_{ij}\, r_1^{\,i}\, r_2^{\,j} \tag{7.3}$$

If the metric $g_{ij}$ is symmetric in $i$ and $j$, then the space is commutative, i.e.

$$d ij = d ji \tag{7.4}$$

If $g_{ij}$ is not a symmetric matrix, then

$$d_{ij} \neq d_{ji}$$

$$\tag{7.5}$$

In flat space the components of the metric $g_{ij}$ are constants in a suitable co-ordinate system. But in curved space derivatives of the metric give the curvature of the space, which in turn give the acceleration or force acting on any particle. In fact, the dynamical equation of motion in a curved space is given by optimizing the Lagrangian

$$L = \sqrt{-g_{AB}\frac{dX^A}{dt}\frac{dX^B}{dt}} \tag{7.6}$$

yielding the rather complicated for

$$\frac{d^2 X^A}{dt^2} + \Gamma^A_{BC}\frac{dX^A}{dt}\frac{dX^B}{dt} = 0 \tag{7.7}$$

where the Christoffel symbol is given by

$$\Gamma^A_{\,BC} = (1/2)\, g^{AD}(\, g_{BD,C} + g_{CD,B} - g_{BC,D}) \tag{7.8}$$

where $g_{BD,C}$ means the derivative of $g_{BD}$ with respect to the co-ordinate $x^c$ etc.

This equation is for the motion of a single particle [with coordinate **X**] in the gravitational field given by all other particles in forming the curved space represented by the metric *g*. This is equivalent to an agent interacting with the cumulative field of neighbors, in a fashion different from the spin model formulation, which may be useful in a different context from what we are considering in the present paper.

$g_{AB}$ may contain damping factors for any large component of $X^A$. The situation is similar to of Fisher-Rao (Rao 1945) metrics related to probability distribution functions (pdf). If the pdf is parameterized by the vector $a^A$, then

$$ds^2 = g_{AB}\, da^A\, da^B \tag{7.9}$$

where

$$g_{AB} = \frac{d^2 \log p}{da^A\, da^B} \tag{7.10}$$

In a social/psychological scenario, changes are meaningful almost always only in terms of fractions (percentage growth, literacy rate, employment, mortality etc.). Logarithmic quantification is relevant in information theory; for example, it has extensive use in entropy. The perceived value of a quantity in a social context may be represented more conveniently as a logarithm, i.e. if we define

$$Y^A = \log[X^A] \tag{7.11}$$

where $X^A$ is the original definition of a quantity [e.g. wealth in terms of actual amount of money, and $Y^A$ may be the mathematically more relevant representation [utility, ability to give satisfaction].

# 8. SIMULATION RESULTS

We show two simple simulations in this section to show the range of behaviors that can arise from such interconnected (spin-like) arrays with varying weights (couplings). For simplicity we choose only three variables and two agents in the simulation. More complicated behavioral patterns where clusters of agents are involved and the expression of variables may change to give rise to interesting social emotions and patterns will be dealt qualitatively later in the paper. Our simple simulation with two agents with different preference, while also connected with the environment, show how introducing multi-component arrays where interaction take place between both the agents and a single agent's preferences are also allowed to affect his other preferences. They also display interesting dynamics and interactions that can vary over time, hence redefining strengths of friendships or ill feelings as the variables affect one another.

$$H = -\sum_{i,k,a,b} A_{ik}^{ab} s_i^a s_k^b - \sum_{i,a,b} B_i^{ab} s_i^a s_i^b - \sum_{i,a} h_i^a s_i^a$$

(8.1)

The $A$ and $B$ matrices are, for convenience, separated forms of the $J$ matrices of coupling given in Eq. 2.1. In $A$ we have mutual interactions among agents, and in $B$ we have interactions among different attributes of the same agent. The interaction with the environment is again represented by the third term.

There are nine $A$ matrices coupling various variables between different agents, three $B$ matrices coupling to different attributes of the same agent, and one $h$ matrix coupling each agent with the environment. Simulations showed irregular behavior and punctured equilibrium depending on how random the initial interaction matrices are taken to be. In this simulation the randomness was chosen to be large to show the possible array of dynamics. However, it is possible that two agents placed in the same society will show more similarities and fewer differences with oscillations showing at larger time scales than our simulation.

Three plots are presented below:

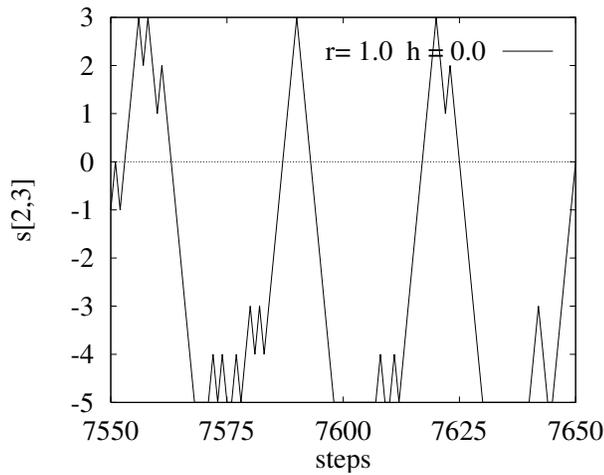

Fig 1. Short period oscillations of $s^2_3$ [ attribute 2 of agent 3], modulating longer period gradual reversals. Compared to the fixed A and B matrices of couplings, there is also a random noise with a relative scale factor = 0.33, external orienting field strength h= 0.

In Fig. 1 the absence of any external field indicates that the agents are not influenced by the environment (which in real life will not be the case and the connection to a common environment will dampen many dissimilarities that would have otherwise interacted with each other and oscillated). The ordinate indicates the changes in the attribute/preference number 2 of agent number 3, which, since we are using a spin model can only vary between -5 and + 5. It is interesting to note that (a) despite the large randomness enforced, the behavior of changes is fairly regular, (b) the state stays more near the -5 extreme [ extreme dislike or liking, depending on how we define the signs] than at the other end. We should, however, remember, that this plot results from a specific choice of the coupling mutual coupling matrices *A* and self-coupling matrices *B*. So, we have a situation which makes an agent change his preferences drastically on account of interactions with others and with the environment and even some noise,

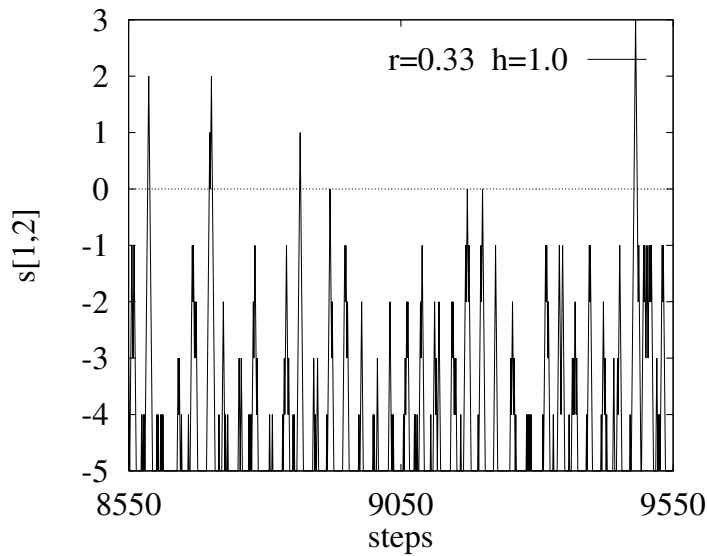

Fig. 2: Near chaotic long-term behavior for $s^1_2$ [ attribute 1 of agent 2], for random noise scale = 0.3, external field h= 1.0.

The results of Fig. 2 can be construed as the same agents with an equal amount of dissimilarity trying to affect and convert each other while also being influenced by a steady environment [the A matrices coupling the different agents are formed in such a way as to quantify the effects of the dissimilar preferences in that fashion]. It is interesting to note that since the environment is constant, it will affect the opposite attributes in the agents in opposite ways. In addition, in real life components of the environment will couple to only specific attributes and not to all of them in the same way. For example, the weather would be more related to the clothing habit of an agent and will strongly couple to it to flip the clothing style of two agents in the same direction. To simulate that possibility we have taken h to be a matrix that has different couplings to different variables. However, it is likely that similar variables in different agents will have similar h values.

In Fig. 3 we see that a certain attribute of a particular agent may show punctuated equilibrium, i.e. it remains more or less constant most of the time, but that is interspersed with sudden bursts of non-equilibrium fluctuations.

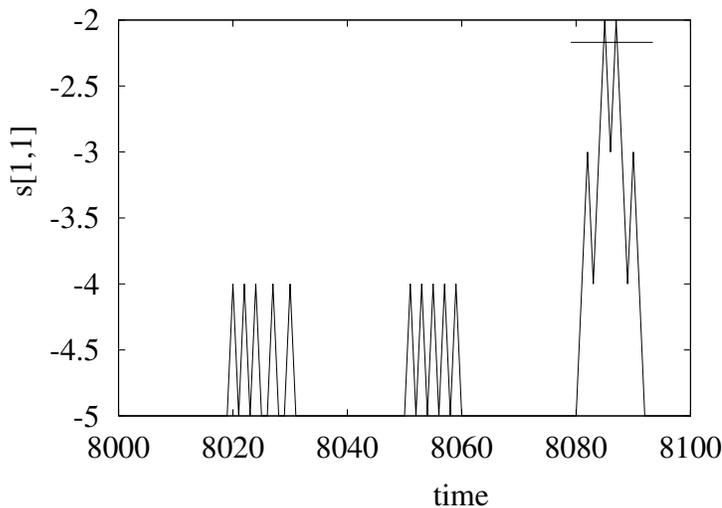
Fig. 3: Punctuated equilibrium seen in attribute 1 of agent 1.

The concept of flipping energy was mentioned in section 5. Part of this flipping energy or resistance to flip is provided by matrix *B* where the variables in the same agent are correlated. Part of it is also provided by the last term where the variable is coupled to the environment *h,* so that a second agent trying to flip the variable is resisted by the coupling with the environment.

## 9. AN AGENT IN A GROUP PSYCHOLOGICAL PERSPECTIVE

The simulations in the section above showed some very simple cases with only two agents and the correlation matrices were not modeled after any specific scenario to show the possible dynamics using this model. When more than two agents and a number of variables with certain correlations and fixed states are in play, interesting social properties may arise from this model when the variable-states and constraints approach different limits. Most of these properties involve a large number of agents and variables and a number of steps. In this paper, we describe how certain social emotions may arise by considering our model in limit cases by analyzing by words. More precise equations with equilibrium points and detailed complicated simulations will be presented in follow-up papers.

### 9.1 Unhappiness/Frustration

The economics of happiness is now an established concept (Kehneman

1999) and it ha also been seen that surveys of people claiming happiness are often inconsistent with wealth (Marks et al 2006)

We further extend it in terms of an agent's state in our social cluster.

An agent might experience cost of inefficiency or non-optimum connection in the form of unhappiness.

1. An individual agent with locally excited preference states to minimize total interaction cost will be unhappy. A forced excited local state might be a result of too many close neighbors in the interaction space where the agent has a fixed ground state variable that is conflicting. An agent with too many close neighbors with conflicting states but not with the bargaining power to leave will remain in a locally excited unhappy state.

2. An agent's sense of insecurity as a result of too much distance from other agents in the interaction space leading to lack of knowledge about the other agents' states who themselves might be closer to him in the interaction space or the security of collaboration especially when needed when interacting macroscopically with the environment.

An agent in need of collaboration will try to be a close neighbor of as many agents as permitted in the interaction space. He will try to influence and flip these agents if necessary, but an agent not in need of collaboration will tend to minimize interactions, especially with agents who might have conflicting variables as those in need of collaboration might try to flip the variables to "convert" him, which might lead to a waste of energy.

Optimally an agent will tend to minimize his own excitation of states by withholding information about variables not required for interaction or collaboration regarding other variables while showing only a certain variable where alignment for that particular collaboration is needed.

However, a second agent who has more information about the ideally hidden states will have higher bargaining power over the second agent.

## 9.2 Trust and Social Norms

We have already mentioned that networks may be created when agents choose to associate with others who are similar to themselves in some great respect (Lazarsfeld Merton 1944). Recent papers have tried to identify an agent's communities depending on genetic and cultural clusters (Gintis 2003a; Kaoki and Feldman). Our self-similarity based model is able to explain the findings of these papers.

The array we assign to each agent, containing weighted variables of both genetic and cultural/belief type, is able to explain an agent's allegiance towards clusters of different types. The changeable weight terms explain the how although an agent's identity contains multiple variables how one type may appear to be more important than another at a certain time given specific conditions. The interactions among peers and kin may modify the weights of the variables and one may appear as more important than another variable in an agent's realization of his identity at a given instance. Hence, with the aid of modified weights, we can define splits within a supposedly coherent cluster and situations like betrayal. Sudden rises of short-term identities that lead to long-term effects can also be explained.  A long-term effect of group sacrifice for group utility could suddenly give rise to splits based on separate contrasting terms in the array over a short period and lead to strong emotional surges.

Gintis (2003a) observes how an agent's internalization of social norms depends on vertical, oblique and horizontal transmission of knowledge.  An agent's trust in the knowledge transmitted to him depends on the frequency of the information and the source's reliability.  The transmission of knowledge in forming norms has been shown to be evolution wise stable (Gintis 2003a).  If an agent had to acquire information about the environment from the scratch, civilization would be static as the growth of civilization is based on accumulated knowledge and the transmission of internalization is based on passing of accumulated stable or workable strategies as inherited from sources that can be trusted.  Studies (Henrich et al 2001a Henrich et al 2001b)  show that if agents from different cultures are presented with an ultimatum, they do not all come up with the same rational strategy, but behave according to social norms they had acquired. A norm evolving as a stable strategy within a cluster is often taken for granted by an agent who has remained in the influence of that cluster. Instead of reinventing strategies each time an agent is presented with a problem, he would rather take many of these social norms for granted and act accordingly even though now not in context with the same social cluster.  In our model, the internalization of a norm may be attributed to including that norm as part of the agent's identity that is taken for granted unless further interactions are able to flip it to a different state.

Our model can show how the trust in all three types of transmission can be established based on the agent's realization of similarity of all three types of variables.  Parents pass norms to their off springs as a propagation of strategies associated to their identities.  However, conflicting norms from parents inherited simultaneously may produce conflicting internalizations taken for granted as trustworthy strategies passed down in generations.  The transmission of horizontal internalization is more complicated as the similarity of agents passing the norms remains more ambiguous and subject to credibility test.  However, as the number of agents in a network increases, the passing of internalizations within the network has the effect of the same information being passed to the same agent with a much higher frequency and the credibility of the information increases as more of the agents one agent is connected to in a network deliver the same information. The transmission of norms in this mode often creates fads, whims and mass mobilization in a network.  Since the similarities of the agents one is connected to in this manner is added up with the same piece of information while the differences average out if the network is based on some broad similar characteristics while other characteristics among agents are distributed randomly like in a Gaussian.

We can mathematically express this behavior of effects from agents within the same cultural cluster having more effects on one as follows:

Let us assume that an agent is connected to N agents who form a cluster based on their similar traits in variables A, B and C while the agents are randomly distributed in variables D, E and F.  With low N's differences in D, E and F may not cancel out and the credibility of any information passed by the neighbors will not have an effect of being accepted for granted.  A rational agent will tend not to accept information gained in such a form without verification and although there might be a bias formed towards the information, it will not be taken for granted and the information will remain as a low weight variable in an agent's preferences or motivators.  However, as N increases, the piece of information is related with the same states in A, B and C N times while the differences in states D, E and F form a Gaussian distribution with an average canceling effect.  Information gained in this manner will develop a high weight and credibility and may prompt an agent to identify with the new piece of information as an internalization that he takes for granted.  This type of propagation of horizontal information may create mass panic as the same information is fed along the connected network with high frequency.  In a modern society, the media might be able to produce a similar effect as the same piece of information is fed to N connected agents who again interact with one another with the information.  In the absence of other sources contradicting the information, this type of information transmission may produce a group of people who are highly biased towards a preference or news.

Elsewhere (Shafee and Steingo 2008) we discuss how momentum in vertical knowledge and internalization and the weight associated with naturally because of genetic similarity often connected with long-term interaction and childrearing often acts as resistance to changing internalizations after sudden changes in social structures take place.

## 9.3 Less than rationality and Altruism

Recent papers (e.g.  Bowles and Gintis 2003; Gintis 2003a) show that rationality at the individual level is often not the correct explanation of how human beings behave.  The stated papers show how pro-social behaviors like altruism may be programmed as an evolution-wise stable strategy necessary for the fitness of a cluster.  It has been shown (Gintis 2003a) that a few agents who are altruists within a cluster as a matter of having altruism a subjective selection with no individual fitness value can be advantageous to a cluster and may contribute to the cluster's stability.  The authors explain this evolution-wise strategic behavior as prosocial emotions programmed.  Our model explains how this prosocial emotion can actually be created based on self-similarity.  The basis of pro-social behaviors was upon the fact that altruism often takes place when reciprocity is not expected in the last step of a game or with strangers one is not expected to interact again.  The authors try to explain these acts as strategies that were stable within a cluster simply because they contributed to the longevity of clusters especially in times of possibilities of threats to the cluster when agents are likely to escape.  However, the authors do not take into account how societies also dissolve when there are no apparent threats but internal strife fragments clusters allowing "outsiders" to take advantage.  In another paper (Shafee and Steingo 2008) we explain how our model, based on self-similarity and varying weights of the variables associated with the concept of self, is able to take into account phenomena of internal conflicts and "selling out" one's cluster by the same agent who once might have taken "altruistic risks" for the formation of the cluster. as well as behaviors of one-time altruism.  Examples are ample in history when a leader who had once led to a country's freedom at one's risk and by chanting nationalism had later risen to power and were later blamed to be corrupt against the same people for whom he had once risked his life (e.g. see Ellison 1988).

Here, we discuss how prosocial emotions like altruism as have been studied to exist may arise from our model in a manner that allows the "feelings: to undergo changes given different situations.

The acts of selflessness can be explained in terms of a calculation of similarity and the perpetuation of the idea of "self" in an apparent game when the agent is actually perpetuating the "most" of self similarity in within the longest time span after doing a rough calculation in risks. When risks of the idea of "self" perishing are the most, feelings like altruism may arise if some parts of self similarity may be protected in this manner. In other extreme cases, under equally perilous conditions, one may desert others in the most inhuman manner, although these acts might later be followed by situations of guilt.

In a recent paper (Bowles and Gintis, 2003) it was shown that oblique transmission of norms is necessary for altruism in an independent model that does not take self-similarity into account. We can show that this is more plausible in societies where the sharing of ancestors among random members of the same cluster would create a more homogenous set of social norms, so that an oblique transmission would also be implying transmission of norms from agents with whom one might share ancestry. Hence, an oblique transformation would imply the transformation of knowledge to agents with whom one shares cultural norms and also genetic variables. So an older member of the cluster, when transmitting norms to a random member, would share both cultural norms bolstered by partial overlap in genetics and the credibility in the form of passing norms to agents with whom he shares similarity in both genetic and cultural norms. Instances of oblique transmission in the form of a random older member of a cluster holding a respectable position among the younger can be shown to be more evident among ancient cultures where a high degree of sharing of norms and genetic markers in a braided through generations. It is in societies of this form too where larger degrees of "selfless rituals" and sacrifices can be observed. Hence, models that claim that oblique transmission of norms are needed for instances like altruism can also be explained in our model.

We also observe the effect fluctuation and faking agents in such homogenous societies where trust is taken for granted. Although stable in norms and efficient in the sense that information is easily taken for granted, saving one the time and energy to reinvent strategies, a cluster where homogeneity s achieved in this form may perform poorly in response to shocks and fluctuations because of mutations, a few faking "outsider" agents who are mistaken to be "insiders" and hence trustworthy. However, the information is again a function of the size of the cluster and an outsider can fake easily as an insider where a cluster is huge and homogenous because of each pair of parents having a large number of progeny but sharing a homogenous set of norms because of the proximity of closeness of the cluster.

Emotions such as guilt and shame have been shown to play roles in how people behave even in some instances where there is no evident punishment for maximizing one's utility (Gintis 2003b). Rather than stating that the emotions are simply evolutionary built in, the phenomena can be easily explained in terms our self-similarity based model. If the notion of self is extended to include self-similar variables, then as an agent tries to maximize the notion of identity associated with his own expressed array, similar components in other agents are also taken into account. As one tries to maximize utilities based on individual differences, after a certain amount of maximization, the concepts of similarity are expressed as similarities with others corrected for dissimilarities start to express themselves and the maximization process for an individual is inhibited. Similar situations may arise in cases of wars and fractured societies when differences become acute by the process of accumulation and suddenly cause actions that leave one of the parties in an extreme state. The aftereffects of such extreme situations where sudden feelings of strong difference give rise to rapid actions leading to destructions and murder often include feelings of guilt. These common situations have been portrayed successfully in literature (e.g. Nicholas 1988).

This feeling is some times coped with by selecting agents with the same differences and offering them extra help. However, situations abound when the strong or threatening members of the group with differences are destroyed and selected members from the weak are shown generosity to balance the senses of similarity and differences (see for example Wallace 1998 for a story of adopting slaves without abolishing slavery).

## 9.4 The Turchin Model

The Turchin model (Turchin 1990) can be examined under our model. According to the Turchin model, once a population size exceeds an optimal value, wars and splitting start to take place. In a recent paper (Shafee and Steingo 2008) we discuss the skill hierarchy in a complex social network and how levels in that hierarchy can accommodate only a certain number of agents and how the support of that hierarchy pyramid most have a degree of similarity.

As the population size increases, the preferences get split into several subgroups. This might lead to overflowing of a certain skill level with contradictory preferences or a certain preference being pushed to the lowest hierarchy level while members of a contrasting preference occupy the topmost levels. In such cases, after the population reaches a certain degree, subclusters are formed with specific interest and conflicting preferences arise in a macroscopic level (Shafee and Steingo 2008). The same population is split into two once the population exceeds what the resource can support (Gumerman et al 2002). However, overpopulation leading to scarcity can divide a population into conflicting factions (Bhavanani and Backer 1999), even if the population could have been supported given ample resources. This may be because, as resources get scarce, an agent adheres to more self-similar agents in the hope of perpetuating his identity by forming larger factions that are able to cooperate for larger changes or fights against others. However, as these factions are formed, violence may arise to dissipate the scarce resource.

We note with interest that existing models using spin glass (e.g. Sinha and Raghavendra 2004) that rely on the phase transitions using one or two variables in an otherwise static society can easily be retrieved as a lower dimensional projection of our model. If we fix all the other variables in agents as the same for simplicity in a fairly homogenous society where the highly weighted terms are similar, the propagation of one or two emerging variables like the spread of rumors or diseases can be retrieved easily. Hence our model maps back to the known models that can often actually lead to nice phase transitions as the dimension of differing variables is lowered, but give rise to more interesting behavior and often chaos once the number of dissimilar variables is increased.

**9.5 Number, Accumulated dissimilarity, Preservation of Similarity: From Friends to Enemies: Altruism to the "Selfish gene" and Oscillations in between**

We briefly show how changing the total number of slightly dissimilar agents placed in a cluster affects the behaviors of the agents. Situations of peril subject to the total number of agents with dissimilarity variables in random places, and given the constraint of maximization of similarity with an individual's basic array starting with the highest weighted term may produce extreme cases of altruism or extreme selfishness for kinship as limit points and the vast spectrum in between. Also an agent can shift between being friends to enemies with the same group of agents as numbers are allowed to change.

These situations will be modeled mathematically in a follow-up paper. We provide word arguments here.

As a group is started with a small number of agents and the total number is allowed to grow, after a threshold is reached the slight dissimilarities among agents add up to produce a vacillating array with an added up chaotic dissimilarity, with each agent being similar in most places but slightly dissimilar in a few places. The total dissimilarity from a large number of agents is random, and chaotic. In such a large group where the dissimilarity Hamming distance between any two random agents is close, but with dissimilarities in different places, the formation of stable groups is tough as with the number going up, total dissimilarities and conflicts of interest add up to create no stable interest. The difficulty of choosing one subgroup over the other prevails, as well as the complexity to form a workable cluster with a functioning group with similar interests but varying skills. In such a situation, the only stable factor leading to the formation of a sub-cluster may be genetic kinship, which remains stable over generations and may form stable patterns of subclusters. However, these sub-clusters will be small and without the complete set of skills required to survive. In order to function as a group, other variables such as incompatible personalities and skills might be forced to become compatible to keep the genetic based cluster steady and functioning. Some of these incompatibilities may be difficult to flip leading to disastrous results. This situation can be somewhat similar to large degrees of ethnic nepotism in slightly diverse high population areas (see e.g. Vanhanen 1999, Ahmed 1991)

As the total population goes up, each of these kin-based subgroups compete for a limited resource. To increase the share of each group, and to enhance the size of the group in order to have a larger fractional share, leads to a higher total population and a more acute severity of the scarcity in a fashion similar to the "Tragedy of the Commons" (Garrett 1968).

At the other end of a spectrum, the following can be observed. A group of self-similar agents functioning as a society may be in peril because of the presence of another group with a larger total dissimilarity due to cultural isolation etc. They can also face extinction because of environmental peril. When the chance of survival of an individual is low, together with the chance of the survival of the cluster, if a large number of members from a similar cluster can be saved at the cost of a member who has a higher chance of dying, total similarity factor of the individual will be increased by behaving altruistically. This similarity factor can be derived from a mixture of factors leading to identity, including kinship, faith and historical bonds.

The case of oscillation from friends to competitors or enemies in the same case, as a situation in between these two extremes can be observed by increasing the total number of agents slowly. A few agents with say a dissimilarity hamming distance of 5, if placed in an environment with odds, may cooperate and form a group of friends despite the differences. However, as the total number of agents is increased, grains of sub-clusters are formed around each of these agents with the new agents with lower difference hamming distance. If they are all in the same environment the once friends may now form their own competing subgroups who are rivals (see e.g. Flint, 2006).

The last note we would like to add is about perception and locality in a global network. It is interesting that an agent interacts with the environment personally at a local level. Hence, his perception and knowledge about the environment in the first degree are dictated by the local environment. However, as agents are connected in a network that spans globally or over a larger span, inter-agent communications are used to form a wider range of knowledge and information that are of mutual interest. However, how the globally acceptable information is contradictory with the conflicting localities and local conditions is interesting to note. Each agent's perception and needs are weights by the environment that directly affects him, which is to the most extent local. Hence, the priorities or weights of an agent are by default influenced by a local environment where inter-agent communications from global networks are also fed and superposed onto it. Hence, the two competing sources of information may at times be even conflicting or may attune the weights of an agent that is not optimal with respect to the other source.

## 10. CONCLUSIONS

We have outlined the possibility of developing concepts and relations related to the evolution of social clusters, analogous to spin systems, and the importance of the concept of the "self" of each agent with variable attributes which may be quantifiable. We have considered the compartmentalization of the entire set of agents into subspaces where mutual influences are significant, and have defined an "influence distance" that may help such clustering. We have shown some interesting analogies between the equations of motion in general relativity and in the evolution of a single agent, i.e. a kind of geometrodynamics in influence space.

Simulations with weight factors for different couplings between agents and their attributes and spin-type flips in either direction from consideration of a utility function in a simple toy system seem to show the possibility of chaos, or at least highly aperiodic behavior, with also the possibility of punctuated equilibrium-like phenomena. It would be interesting if the reverse process of obtaining the *A* and *B* matrices coupling different agents with different attributes from real data can be successfully realized. However, because of the very large number of parameters available, it would probably almost always be necessary to reduce the problem to simpler systems with a manageable set of matrices of links, using assumptions of fuzziness or symmetry or some other consideration.

## Appendix: Overlapping Perception and Utilities

Due to slight variations of the genetic makeup of perceptions, various agents, although connected to the same environment, may be able to perceive it slightly differently. Hence, they will have slightly different ideas about the world. This is analogous to being connected to a shadow of a higher dimensional world where all agents are connected to highly overlapping but slightly different shadows although interacting with one shadow changes the entire higher dimensional world and hence alters another shadow as well. So two agents connected to slightly different shadows, which cannot perceive each other's world completely, cause changes to the other agent's world by interacting with his own world. Although the first agent will conceive his own interactions with his own world completely rational, he might find the interaction of the second agent with respect to the first world irrational, even though the second agent himself will find interactions with his own world perfectly rational.

**Acknowledgement:** The author would like to thank Prof Phil Anderson and Prof Michael Fisher for going over the preliminary draft. She would like to thank Prof Douglas White for providing valuable feedback and suggestions. She would also like to thank Andrew Tan for discussions and his continued moral support.